\documentclass{PoS}
\usepackage{amsmath}
\usepackage{amssymb}
\usepackage{xfrac}
\usepackage{soul}

\newcommand{\LL}{\Lambda^2}

\newcommand{\calO}{\mathcal{O}}
\newcommand{\dd}{\text{d}}

\newcommand{\als}{\alpha_S}

\newcommand{\pt}{$p_T\ $}

\title{BSM effects on the Higgs transverse-momentum spectrum in an EFT approach}

\ShortTitle{BSM effects on the Higgs transverse-momentum spectrum in an EFT approach}

\author{Massimiliano Grazzini$^a$\footnote{On leave of absence from INFN, Sezione di Firenze, Sesto Fiorentino, Florence, Italy.}, \speaker{Agnieszka Ilnicka}$^{ab}$, Michael Spira$^c$, Marius Wiesemann$^a$\\
$^a$ Physics Institute, University of Z\"urich,\\
\hspace{0.23cm}Winterthurerstrasse 190, CH-8057 Z\"urich\\
$^b$ Physics Department, ETH Z\"urich,\\
\hspace{0.23cm}Otto-Stern-Weg 5, CH-8093 Z\"urich\\
$^c$ Paul Scherrer Institute,\\
\hspace{0.23cm}CH-5232 Villigen PSI\\
 E-mail: \email{grazzini@physik.uzh.ch}, \email{ailnicka@physik.uzh.ch}, \email{michael.spira@psi.ch}, \email{mariusw@physik.uzh.ch}}

\abstract{
Effective Field Theories offer a consistent bottom-up approach to parametrise small deviations from Standard Model predictions.
In this work we report on the application of the Effective Field Theory to shed light on effects from high-scale physics beyond 
the Standard Model on the Higgs transverse-momentum spectrum. 
The Standard Model prediction for the transverse-momentum distribution in Higgs boson production through gluon fusion
is augmented by three new dimension-six operators, implying the modification of the top and bottom Yukawa couplings, 
and the inclusion of a point-like Higgs-gluon coupling. 
We present resummed transverse-momentum spectra including these operators at NLO+NLL accuracy and study their 
effects on the shape of the distribution. The proper parametrization of such effects becomes increasingly important for 
experimental analyses in Run II of the LHC.}

\FullConference{The European Physical Society Conference on High Energy Physics\\
		22--29 July 2015\\
		Vienna, Austria}

\begin{document}

\section{Introduction}

The scalar resonance discovered by ATLAS and CMS at the LHC in 2012 \cite{ATLASdisc,CMSdisc} closely resembles the Higgs boson postulated in the Standard Model (SM).
The SM, however, is not able to explain the existence of dark matter,
the matter-antimatter asymmetry and the relatively low scale of electroweak symmetry breaking (hierarchy problem).
Many theories beyond the SM (BSM) addressing the above issues have been developed, which manifest different patterns in the scalar sector and in the Higgs boson properties. As no strict argument exists for the discovery of
new physics at the TeV scale, it is possible that new physics effects are accessible
only by measuring small deviations from SM predictions. A consistent way to parametrise these deviations is offered by 
the Effective Field Theory (EFT), in which the unknown high-scale fields are integrated out leaving an infinite ladder of 
higher-dimensional operators (dim$>4$), with a well-defined hierarchy. The EFT can thus be used to build a bottom-up 
approach in which the usual dimension-four operators in the SM are augmented by 
leading (dimension-six) operators. Experimental data can be employed
to fix the values of the so-called Wilson coefficients in front of these operators.
A matching to the EFT allows then for the translation of the Wilson coefficients into bounds on the 
physical parameters of new physics models. 
The full set of dimension-six 
\cite{dim61,dim62} and dimension-seven \cite{dim7} deformations of the SM can be classified by 59 and 20 operators, 
respectively. The precision observables measured at LEP and the Tevatron put bounds on many of the Wilson 
coefficients, some even at the per-mille level \cite{fits1,fits2,fits3}. However, several operators involving the 
Higgs field are still essentially unbounded. In the following, we therefore develop a strategy to determine such bounds 
by using the transverse-momentum ($p_T$) spectrum of the Higgs boson.

\section{Transverse-momentum spectrum}

Kinematical distributions provide an important handle on the determination of Higgs properties. 
Among the most important observables in this respect is the Higgs transverse-momentum distribution, that will be measured 
with high precision in Run II of the LHC. First results from the LHC Run I were presented by the ATLAS collaboration 
in the $2\gamma$ and four-lepton final states \cite{atlas1,atlas2} and by the CMS collaboration in the $2\gamma$ final 
state \cite{CMSpt}, leaving still a significant amount of arbitrariness for the precise form of the distribution.
The $p_T$ spectrum provides more information than the total cross section, which is just one number: 
The shape, the position of the 
maximum and the normalisation allow us to disentangle effects that remain hidden in the total rates. For example, it 
is the simplest measurement to shed light on the nature of the Higgs coupling to gluons. 
The fact that the Higgs is a scalar, gives an additional simplification in the modeling of the Higgs 
$p_T$-spectrum, due to the factorization of production and decay in the narrow-width approximation.

The most important Higgs production channel at the LHC is gluon fusion, which, despite being a loop-induced process, is highly 
enhanced by the dominance of the gluon densities \cite{ggfus}. Therefore, we will concentrate on the spectra 
obtained in this production channel. In the past years a significant amount of work has been done to improve the precision of the 
calculations of the Higgs $p_T$ spectrum. The first results at the lowest order (${\cal O}(\als^3)$) were known since long time 
\cite{ptLO1,ptLO2}. It took nearly ten years until the ${\cal O}(\als^4)$ corrections were computed 
\cite{ptNLO0,ptNLO1,ptNLO2,ptNLO3}. These were carried out in the heavy-top limit (HTL), i.e. $m_{t}^2 \gg M^2_H, p^2_{TH}$.\footnote{Finite top-mass effects on the Higgs $p_T$ distribution at ${\cal O}(\als^4)$ were estimated in Refs.\,\cite{Harlander:2012hf,Neumann:2014nha}.} Recently, results on Higgs+jet production at ${\cal O}(\als^5)$ were also obtained in the HTL
\cite{ptNNLO1,ptNNLO2,ptNNLO3}. 

In the low-$p_T$ region ($p_T \ll M_H$), the perturbative expansion is affected by large logarithmic terms of the form $\als^n\ln^m(m_H^2/p_T^2)$, with $1\le m\le 2n$. This results in a singular behaviour of the distribution as $p_T\rightarrow 0$. To cure this problem one needs to resum these terms to all orders in $\als$ \cite{resum1}. The resummation is carried out in impact parameter ($b$) space, and, in particular, we use the formalism of Ref.~\cite{resum2}. The resummed and fixed order results have to be properly matched at intermediate $p_T$ to avoid double counting:
\begin{align}
\left[\frac{\dd\sigma}{\dd p_T^2}\right]_{\text{f.o.}+\text{a.o.}}=
\left[\frac{\dd\sigma}{\dd p_T^2}\right]_{\text{f.o.}}
-\left[\frac{\dd\sigma^{\text{(res)}}}{\dd p_T^2}\right]_{\text{f.o.}}
+\left[\frac{\dd\sigma^{\text{(res)}}}{\dd p_T^2}\right]_{\text{a.o.}}
\end{align}
where f.o.\;corresponds to fixed order, and a.o.\;to all orders calculations. In the formalism of Ref.~\cite{resum2},
a unitarity constraint is enforced, such that
the integral of the $p_T$-spectrum coincides with the corresponding total inclusive cross section computed at fixed order.
More precisely, by performing the resummation at next-to-leading logarithmic accuracy (NLL) and including the fixed order result 
up to $O(\als^3)$ we obtain NLO+NLL accuracy, and the integral of the spectrum is fixed to the NLO total cross section.
Top-and bottom-mass effects can be included in the resummed spectrum along the lines of 
Refs.~\cite{aprmt1,aprmt2}.\footnote{For studies of the resummed $p_T$ spectrum in explicit BSM models see for example Refs.\,\cite{Bagnaschi:2011tu,Harlander:2014uea,Mantler:2015vba}.}

The inclusion of dimension-six and dimension-eight operators in the $p_T$-spectrum has been considered in Refs.~\cite{ptdim61,ptdim62,ptdim63} and \cite{ptdim81,ptdim82}, respectively.
Strategies for extracting information on the Higgs-gluon couplings from the measurements were studied in Ref.~\cite{ptdim63}. Most of the above studies, however, are limited to the high-$p_T$ region of the spectrum, and do not include
small-$p_T$ resummation. In this contribution we present preliminary results for the resummed $p_T$-spectrum at NLO+NLL accuracy, with the inclusion of a set of dimension-six parameters relevant for Higgs boson production.
More details will be presented elsewhere \cite{inprep}.

\section{Effective operators}

\begin{figure}[t]
\begin{center}
\includegraphics[width=0.6\textwidth]{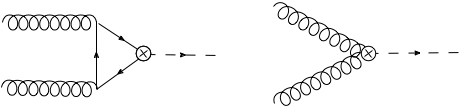}
\caption{Feynman diagrams contributing to the leading-order gluon fusion Higgs particle production with the inclusion of dimension-six operators. In the loop a top or a bottom quark can circulate. }
\label{LOSMEFT}
\end{center}
\end{figure}

In our study, we augment the SM Lagrangian with three new, gauge invariant, dimension-six operators:
\begin{align}
\bar{\calO}_1 = \frac{c_1}{\LL} |H|^2 G^a_{\mu\nu}G^{a,\mu\nu}\,,\quad \bar{\calO}_2 = \frac{c_2}{\LL} |H|^2 \bar{Q_L} H^c u_R + h.c.\,,\quad \bar{\calO}_3 = \frac{c_3}{\LL} |H|^2 \bar{Q_L} H d_R + h.c.
\end{align}
These operators, in the case of single Higgs production, may be expanded as:
\begin{align}
\bar{\calO}_1 &\rightarrow \frac{\alpha_s}{\pi v} c_g h  G^a_{\mu\nu}G^{a,\mu\nu}\,,\\
\bar{\calO}_2 &\rightarrow \frac{m_t}{v} c_{t} h \bar{t} t\,,\\
\bar{\calO}_3 &\rightarrow \frac{m_b}{v} c_{b} h \bar{b} b\, .
\end{align}
The first operator corresponds to a Higgs-gluon contact interaction and the coupling develops the same structure as in 
the heavy-top limit of the SM. The latter two interactions correspond to the modifications of the top and bottom Yukawa couplings. 
In our convention, based on the SILH basis \cite{SILH0,SILH}, we express the Wilson coefficients as factors in the canonically normalized Lagrangian. 
In Figure \ref{LOSMEFT} two representative leading-order diagrams for gluon-induced Higgs boson production 
are shown. 

Note that for this process the total cross section alone does not disentangle the coefficients $c_g$ and $c_t$:
\begin{align}
\label{sigmacgct} \sigma \approx |12 c_g + c_{t}|^2 \sigma_{SM}\ \ (HTL)\,.
\end{align}
Although $c_t$ and $c_b$ may be measured in the $t\bar{t}h$ and  $b\bar{b}h$ production modes\footnote{see Refs.\,\cite{Beenakker:2002nc,Dawson:2002tg} 
and Refs.\,\cite{Dittmaier:2003ej,Dawson:2003kb,Wiesemann:2014ioa,Harlander:2014hya}, respectively, and references therein}
(or $c_b$ through the branching ratio of $h \rightarrow b\bar{b}$)
the total gluon fusion cross section does not give a direct limit on $c_g$.

Our implementation is based on the program HqT \cite{hqt1,hqt2}: a public tool for the calculation of the $p_T$-spectrum of the Higgs boson. The contributions from the dimension-six operators are consistently included to keep NLO+NLL accuracy. The fixed order resulsts are then cross checked with HIGLU \cite{higlu} and HNNLO \cite{hnnlo1,hnnlo2,aprmt2}.

\section{Results}

\begin{figure}[h]
\begin{center}
\includegraphics[width=0.5\textwidth]{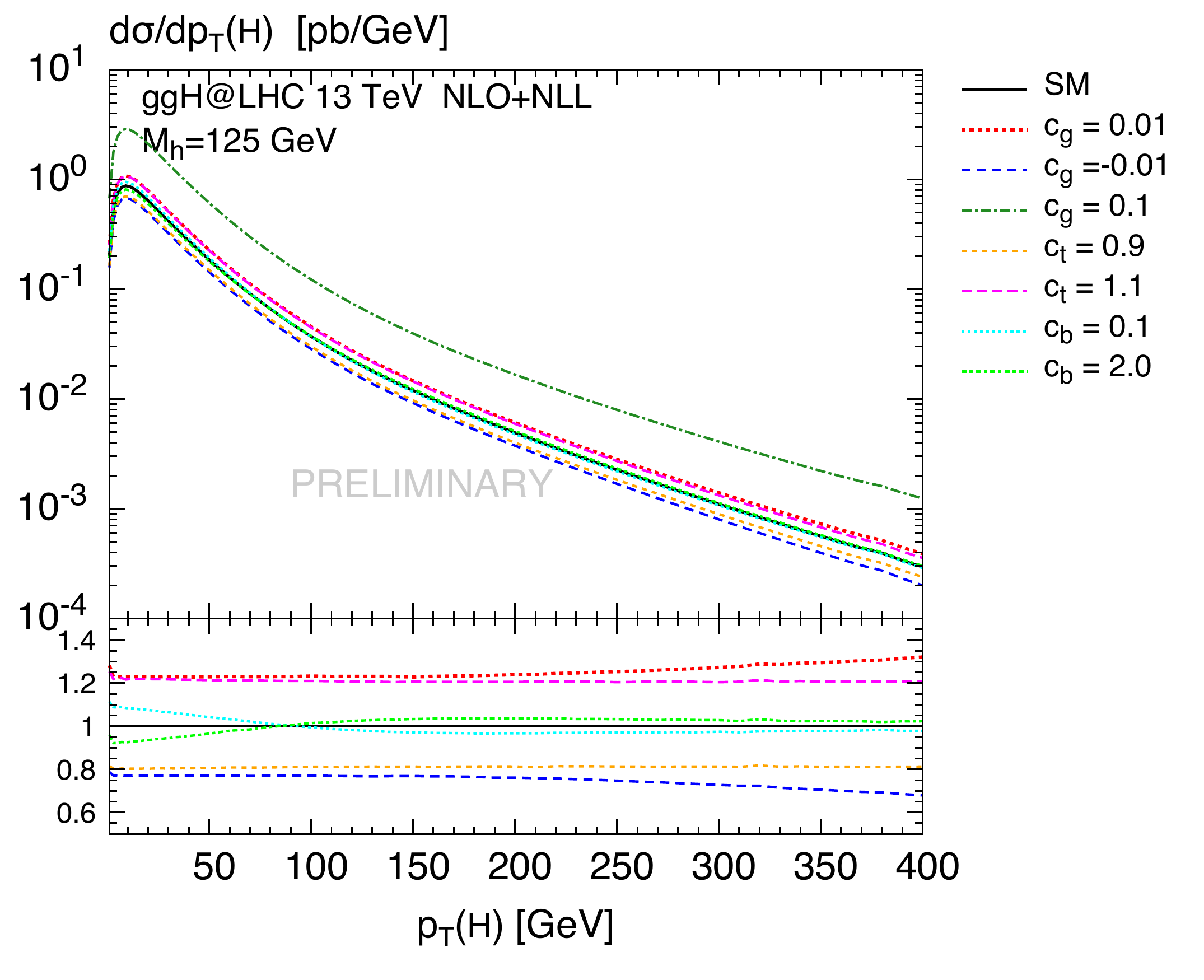}\hspace{-0.2cm}
\includegraphics[width=0.5\textwidth]{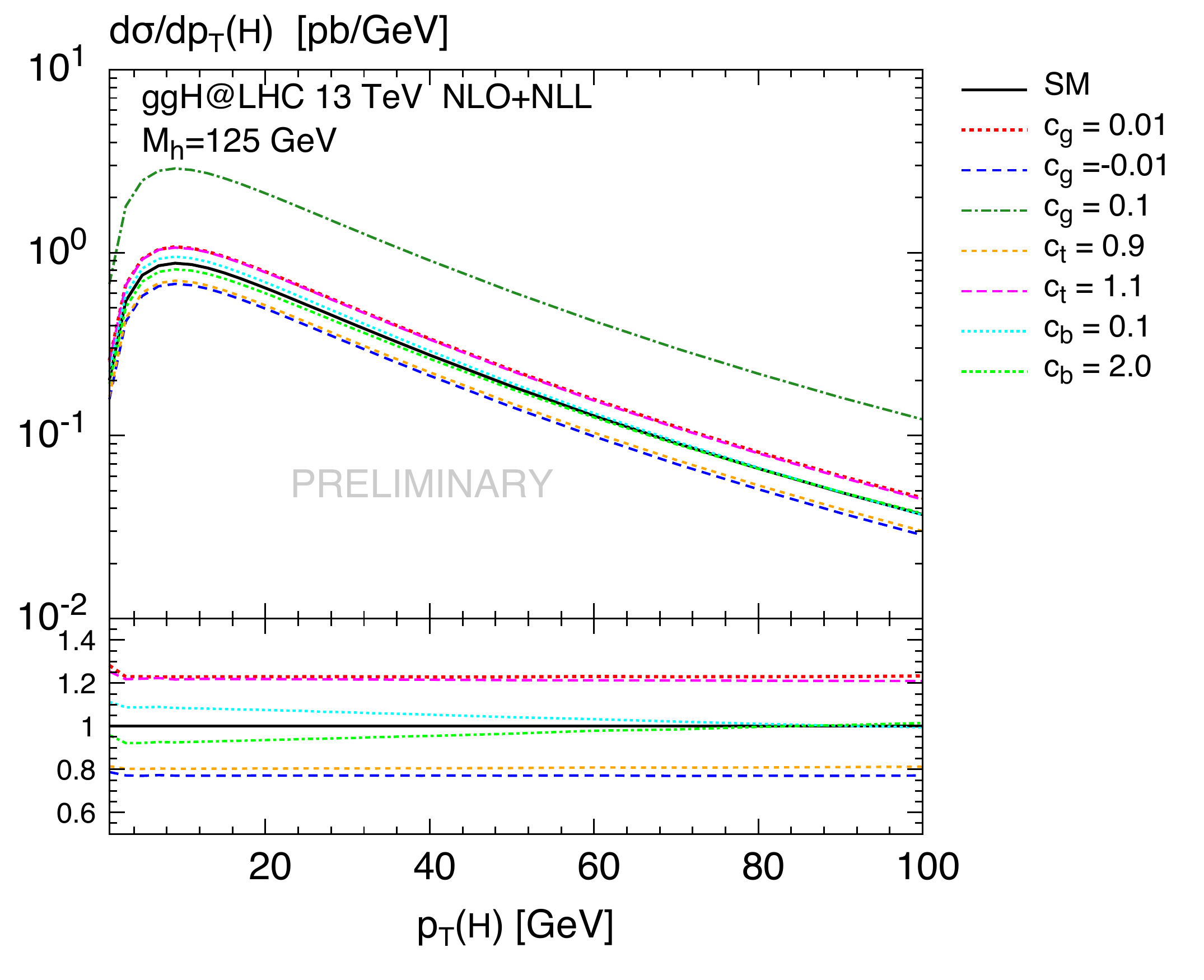}  
\caption{Effects on the Higgs $p_T$ spectrum from separate contributions of the effective operators. The lower panel shows the deviation from the SM spectrum. Right panel is a zoom into the low-$p_T$ region.}
\label{fig:sep}
\end{center}
\end{figure}

In Figure \ref{fig:sep} we present $p_T$ spectra with the contribution from different dimension-six operators separately. 
Values used to modify the SM contributions show that the modification of the bottom Yukawa coupling (by $\sim 90-100 \%$) have 
the smallest impact on the spectrum, mainly through shape changes in the low-$p_T$ region. On the other hand, the point-like 
gluon-Higgs coupling has an impact mainly in the high-$p_T$ region ($\gtrsim 300$ GeV). For small and moderate values of \pt the modifications considered here 
mainly affect the normalization, and to a lesser extent, the shape of the spectrum. 

\begin{figure}[h]
\begin{center}
\hspace{-0.25cm}
\includegraphics[width=0.5\textwidth]{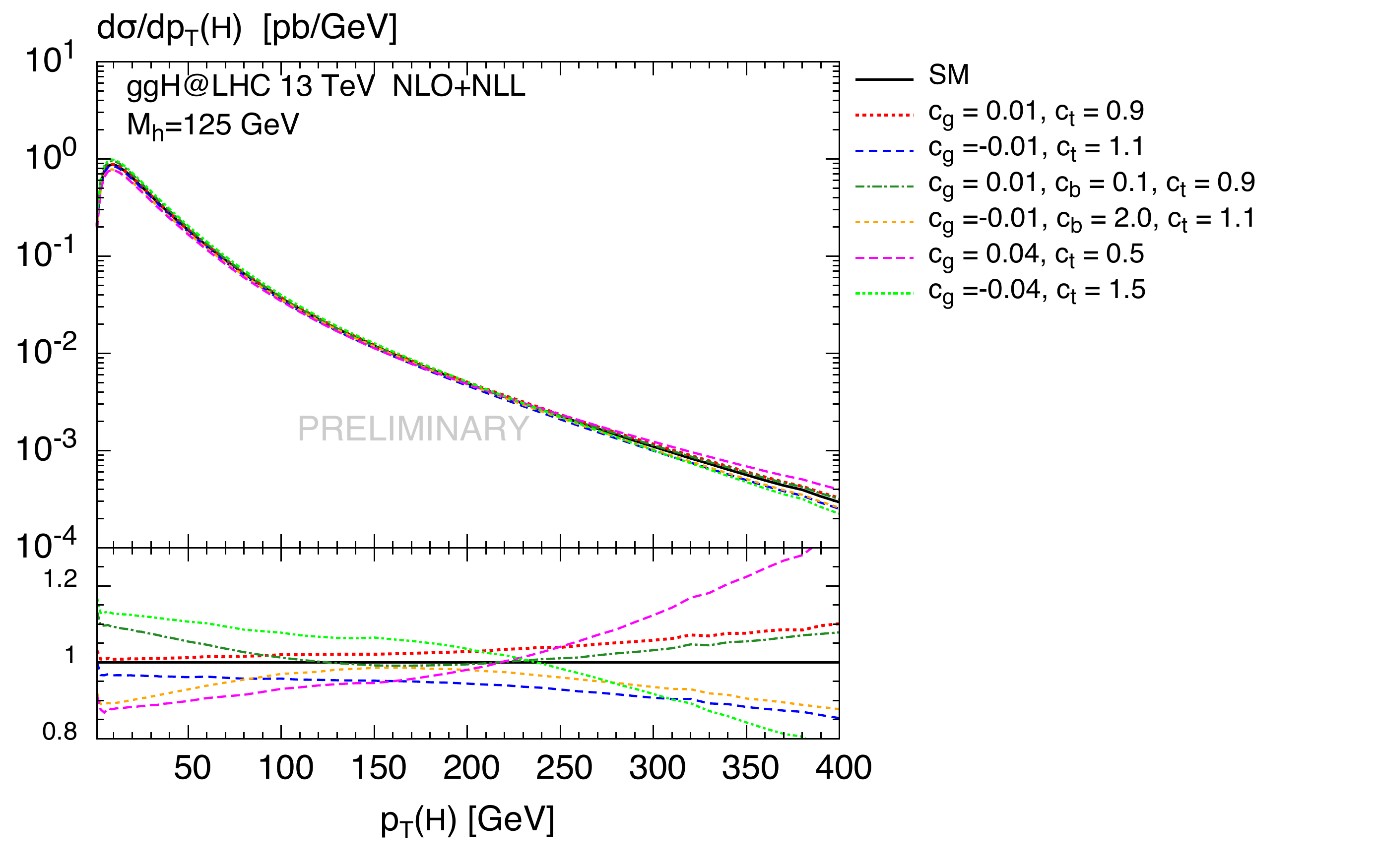} \hspace{-0.15cm}
\includegraphics[width=0.5\textwidth]{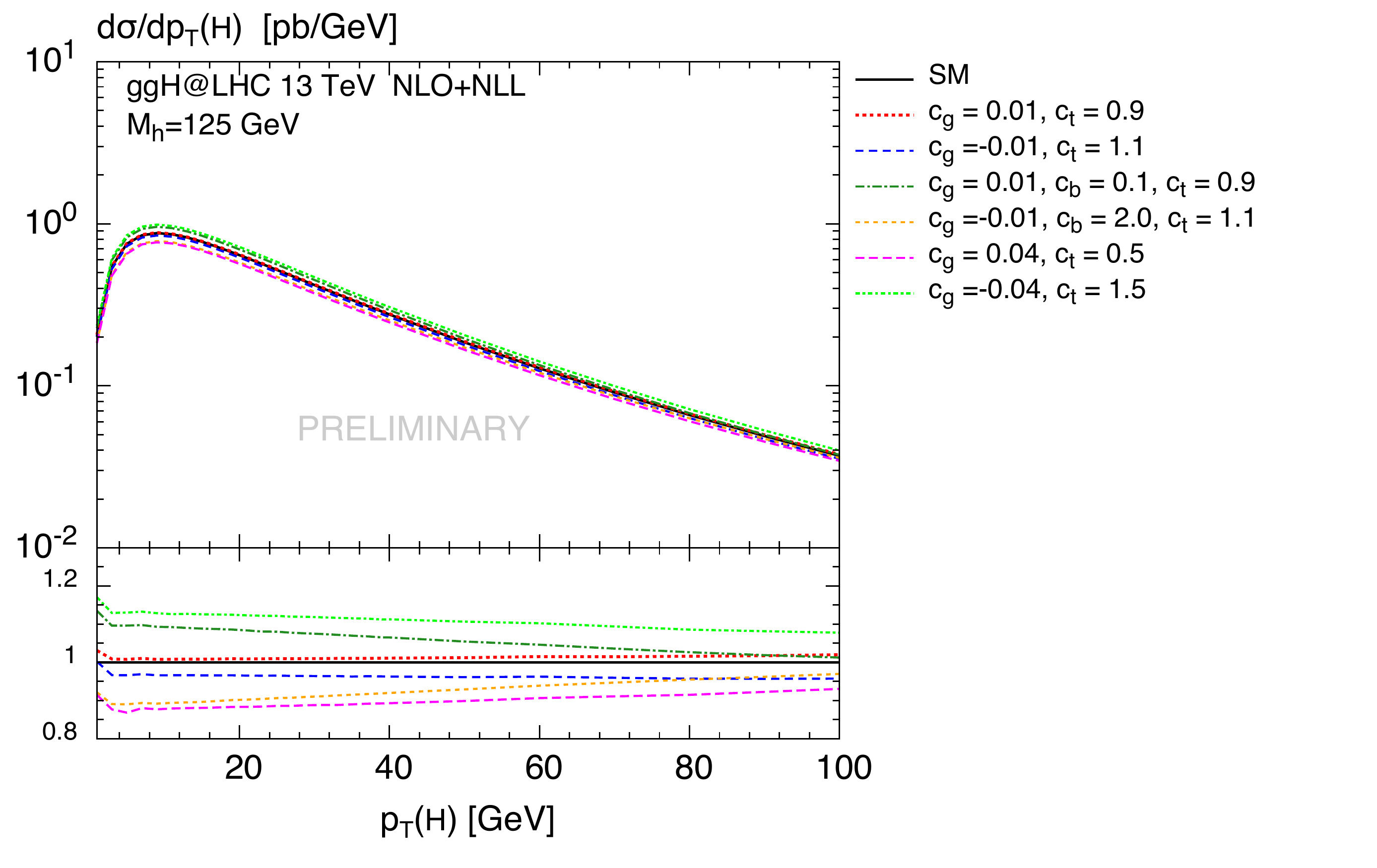} 
\caption{Effects on the Higgs $p_T$ spectrum from mixed contributions of the effective operators. The contributions were chosen so that the total cross section is close to the SM one. The lower panel shows the deviation from the SM spectrum. Right panel is a zoom into the low-$p_T$ region.}
\label{fig:mix}
\end{center}
\end{figure}

Figure \ref{fig:mix} shows the $p_T$ spectra with a simultaneous inclusion of different dimension-six operators. 
The values of the coefficients were chosen such that they produce total cross sections approximately equal to the SM one, 
cf. Eq.~(\ref{sigmacgct}).
We see that large changes in $c_g$ and $c_t$ can compensate each other 
in the total cross section but lead to significant effects on the shape of the 
spectrum, especially in the high-$p_T$ region.  Similarly, a large variation of 
$c_b$ can be compensated by modifications of $c_g$ and $c_t$ in the total rate, 
and then generate sizable distortions of the $p_T$ shape also at small transverse 
momenta. This shows that observables combining low, intermediate and high-$p_T$ 
regions would be able to distinguish between $\bar{\calO}_3$, $\bar{\calO}_2$ and $\bar{\calO}_1$ 
contributions and set bounds on their Wilson coefficients. 

\section{Conclusions}

Effective Field Theory is a powerful tool to parametrise effects of high-scale BSM physics,
which manifest themselves through small deviations from the SM picture, in a model independent way. 
In this formalism the SM Lagrangian is augmented by dimension-six operators. 
We have used this bottom-up approach to model BSM effects on the transverse-momentum spectrum of the Higgs particle. 
Our implementation starts from the ${\cal O}(\als^3)$ result valid at large transverse momenta and supplements 
it with soft-gluon resummation at NLL accuracy. We then include three new dimension-six 
operators: a point-like Higgs-gluon coupling, and modifications of the top and bottom Yukawa interactions. 
Soft-gluon resummation allows us to obtain reliable spectra in the whole range of transverse momenta. 
The contributions from different higher-dimensional operators is visible
in different regions of the spectrum. In particular, a modification of the Higgs-bottom coupling ($\bar{\calO}_3$) 
mainly affects the spectrum at low transverse momenta, while a direct coupling of the Higgs to gluons ($\bar{\calO}_1$) 
changes the tail of the distribution. These results suggest that it will be possible to set
bounds on the corresponding Wilson coefficients once precise experimental measurements of the transverse-momentum
spectrum will be available.

\section*{Acknowledgement}

This work is supported by the 7th Framework Programme of the European Commission through the Initial Training Network HiggsTools PITN-GA-2012-316704.

\bibliographystyle{polonica-mod}
\bibliography{pteft}

\end{document}